\documentclass[a4paper,11pt]{article}
\usepackage{aaskaiid}
\usepackage{orcidlink}

\usepackage{wasysym}
\usepackage{graphicx}

\title{Probing the Fundamental Nature of Particle Dark Matter} 
\ShortTitle{Particle DM at SKAO}

\author[1,2]{Marco Regis\orcidlink{0000-0003-0399-0284}}
\ShortName{Regis et al.} 
\author[3,4]{Aritra Basu\orcidlink{0000-0003-2030-3394}}
\author[5]{Geoff Beck\orcidlink{0000-0003-4916-4914}}
\author[6,7,8]{Gianni Bernardi\orcidlink{0000-0002-0916-7443}}
\author[9]{Paolo Marchegiani\orcidlink{0000-0001-7487-8287}}
\author[10]{Dominik J. Schwarz\orcidlink{0000-0003-2413-0881}}
\author[2]{Marco Taoso\orcidlink{0000-0001-7181-2071}}
\author[1,11,12]{Elisa Todarello\orcidlink{0000-0001-7419-0976}}
\author[13]{Emma Tolley\orcidlink{0000-0002-1027-1213}}
\author[10]{Cora Uhlemann\orcidlink{0000-0001-7831-1579}}

\affiliation[1]{Dipartimento di Fisica, Universit\`{a} di Torino, via P. Giuria 1, I--10125 Torino, Italy}
\affiliation[2]{Istituto Nazionale di Fisica Nucleare, Sezione di Torino, via P. Giuria 1, I--10125 Torino, Italy}
\affiliation[3]{Th\"uringer Landessternwarte, Sternwarte 5, 07778 Tautenburg, Germany}
\affiliation[4]{Max-Planck-Institut f\"ur Radioastronomie, Auf dem H\"ugel 69, 53121 Bonn, Germany}
\affiliation[5]{School of Physics and Centre for Astrophysics, University of the Witwatersrand, Johannesburg, South Africa, WITS-2050}
\affiliation[6]{INAF - Istituto di Radioastronomia, via Gobetti 101, 40129 Bologna, Italy}
\affiliation[7]{Centre for Radio Astronomy Techniques and Technologies (RATT), Department of Physics and Electronics, Rhodes University, Makhanda 6140, South Africa}
\affiliation[8]{South African Radio Astronomy Observatory, Cape Town 7700, South Africa}
\affiliation[9]{INAF-Osservatorio Astronomico di Cagliari, Via della Scienza 5, I-09047 Selargius (CA), Italy}
\affiliation[10]{Fakultät für Physik, Universität Bielefeld, Postfach 100131, 33501 Bielefeld, Germany}
\affiliation[11]{Leinweber Institute for Theoretical Physics, University of California, Berkeley, CA 94720, U.S.A.}
\affiliation[12]{Theoretical Physics Group, Lawrence Berkeley National Laboratory, Berkeley, CA 94720, U.S.A.}
\affiliation[13]{Institute of Physics, Laboratory of Astrophysics, École Polytechnique Fédérale de Lausanne (EPFL), Observatoire de Sauverny, Versoix, 1290, Switzerland}

\abstract{Understanding the fundamental nature of dark matter (DM) is one of the most significant scientific challenges of our time.
A compelling hypothesis is that DM consists of a new, yet-to-be-discovered particle. Among the leading candidates are weakly interacting massive particles (WIMPs) and axion-like particles (ALPs), both of which can be investigated using observations with the SKA telescopes.
In this chapter, we review the search for particle DM through radio observations, summarizing the current state-of-the-art and presenting forecasts for the SKA-Low and SKA-Mid telescopes in the AA4 baseline design.
Radio searches for WIMPs focus on detecting synchrotron radiation originating from the products of DM annihilation using continuum observations. Competitive constraints on sub-TeV WIMPs have already been derived using SKA precursors looking at dwarf galaxies, galaxy clusters, and the Large Magellanic Cloud. We discuss how the superior continuum sensitivity of the SKA telescopes will allow us to progressively close in on the WIMP parameter space.
The ALP signal arises from its decay or conversion into photon(s), which typically consists of a nearly monochromatic signature, and from rotation of polarization angles of photons interacting with ALPs. We demonstrate how the spectral resolution, line sensitivity, and polarimetry of the SKA AA4 telescopes can be leveraged to constrain the ALP-photon coupling.}

\begin{document}
\maketitle

\section{Introduction}
The presence of anomalous gravitational potentials in the Universe has been established through various astronomical and cosmological observations across different scales. Explanations based on modified gravity face challenges from observations of structures at large scales and the description of structure formation through the evolution of matter perturbations~\citep{Pardo:2020epc}.  
These anomalous gravitational potentials could instead arise from a new form of matter, comprising more than 80\% of all matter in the Universe~\citep{planck}. However, the nature of this ``dark matter'' (DM) remains completely unknown.
The Standard Model (SM) of particle physics describes all known fundamental particles and their interactions, and has successfully predicted a wide range of experimental results~\citep{ParticleDataGroup:2024cfk}. Nevertheless, none of these fundamental particles can play the role of DM. Proposed DM candidates within the SM involve bound states formed in the very early Universe, either via phase-transition associated with electro-weak or strong forces, such as QCD balls, or via gravity, such as primordial black holes. However, modeling these bound states is challenging and typically requires a high level of fine tuning.
Several extensions to the SM provide particle candidates for DM. Two particularly compelling classes of candidates are weakly interacting massive particles (WIMP)~\citep{wimps}, and axions~\citep{Sikivie:1983ip}.

WIMPs are neutral particles with mass in the GeV-TeV range and interact weakly with SM particles. Stable WIMPS are predicted by supersymmetry~\citep{wimps} and several other beyond-the-standard-model (BSM) theories~\citep{Bertone:2010zza}.
In the early Universe, WIMPs would have been in thermal equilibrium with the SM, with their relic abundance resulting from a freeze-out mechanism. 
The predicted relic density of thermally produced WIMPs is in the ballpark of the observed DM relic density, a coincidence dubbed the ``WIMP miracle''. The weak interaction implies that there is a (small but finite) probability that WIMPs in DM halos of astrophysical structures can pair-annihilate into detectable species. In Section~\ref{sec:WIMP}, we will describe how to search for WIMP DM using radio waves and the SKAO. This search can be also used to place constraints on decaying DM with mass in the GeV-TeV range.

QCD axions are hypothetical particles introduced to solve the so-called strong CP problem~\citep{Peccei:1977hh,Peccei:1977ur}, namely to explain the extremely small value of the 
neutron's electric dipole moment. 
These pseudo-scalar Nambu-Goldstone bosons, associated with the Peccei-Quinn symmetry breaking, can be produced in the early phase of the Universe, before or after inflation~\citep{Preskill:1982cy,Abbott:1982af,Dine:1982ah}.
In this context, the axion can also be a viable DM candidate, with typical mass $m_a$ in the range of $10^{-6} $ to $ 10^{-3}$~eV. 

Axion-like particles (ALPs) are present in many BSM theories, including string theory models~\citep{arvanitaki2010}. These (pseudo-)scalar particles can have masses as low as zeV and exhibit very weak couplings to the SM, making them good DM candidates. Like QCD axions, ALPs couple to photons through the Lagrangian
term $\mathcal{L} =- \frac{1}{4} g_{a\gamma} a F_{\mu \nu}\tilde{F}^{\mu \nu}$. Unlike QCD axions, the ALP-photon coupling is a free model parameter.
Searches for ALPs involve looking for: their decay into two photons (Section~\ref{sec:ALPdec}), their conversion into photons in the presence of external magnetic fields (Section~\ref{sec:ALPcon}), and the rotation of polarization angles of photons interacting with ALPs (Section~\ref{sec:ALPbir}).

The dark photon is another relevant DM candidate with observational signatures similar to ALPs.
There is increasing interest from the particle physics community in exploring dark sectors, i.e., sectors not charged under the SM gauge groups.
The dark and visible sectors can interact via a kinetic mixing between one visible and one dark Abelian gauge boson.
The latter is referred to as the ``dark photon" and this mixing allows for its potential detection in experiments, see, e.g., \cite{Fabbrichesi:2020wbt}.
In particular, dark photons with mass between $10^{-5}$ and $10^{-8}$ eV can be produced with the correct relic abundance to be good DM candidates, and may be evident in radio observations.

\begin{figure}[h]
    \centering
	\includegraphics[width=0.95\columnwidth]{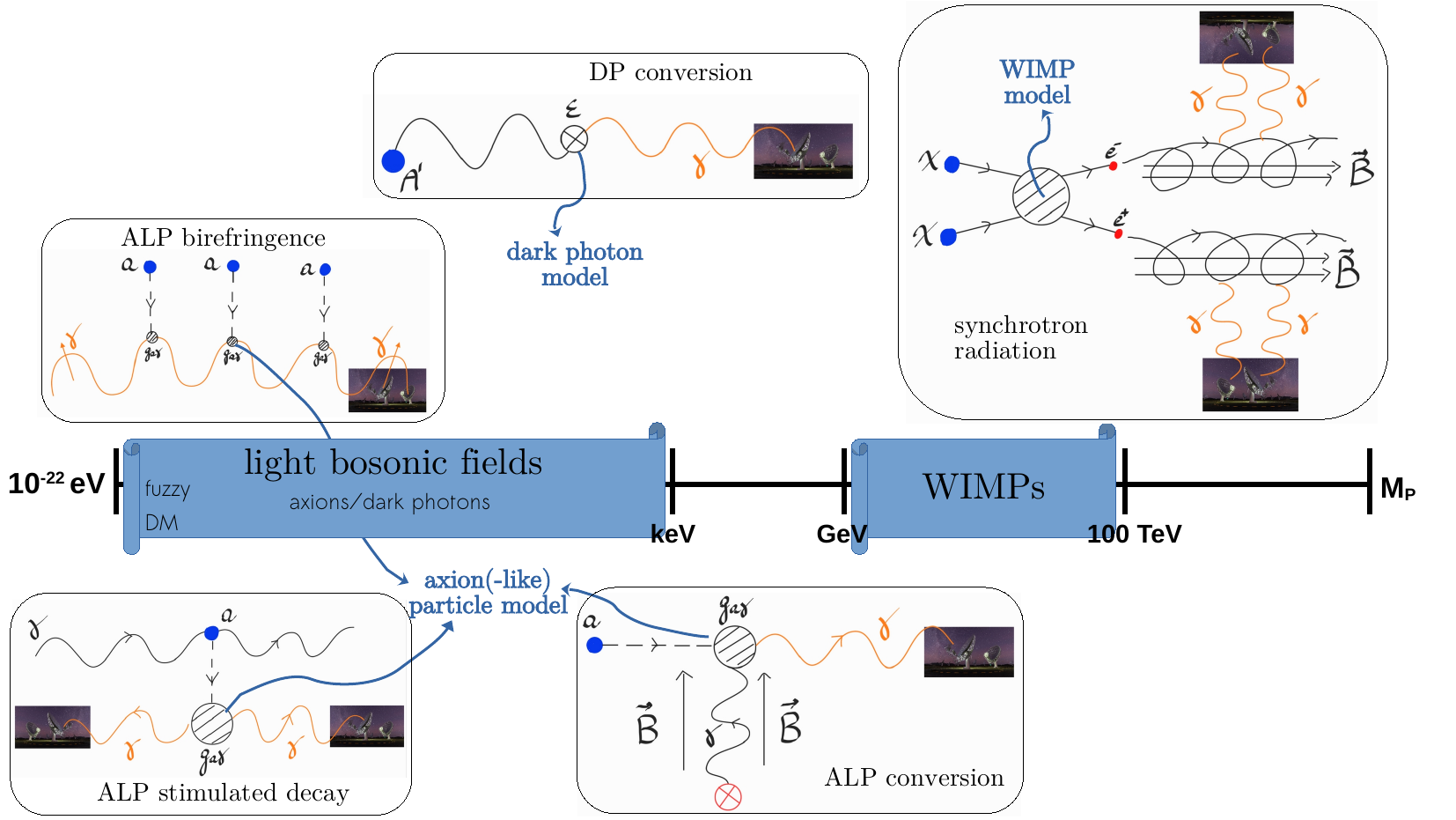}%
    \caption{Cartoon of the different DM candidates and signatures considered in this Chapter.
    }
    \label{fig:cartoon}
\end{figure}

In the following Sections, we outline strategies to detect WIMPs, ALPs, and dark photons using radio observations, presenting projected sensitivities for the SKA telescopes in the AA4 baseline design.
For a quick illustrative overview, see Table~\ref{tab:summary} and Fig.\ref{fig:cartoon}.

\begin{table}[h]
	\centering
	\caption{Summary table for the DM candidates, signatures, types of observation, targets, and observing requirements concerning the capability of the SKAO to constrain particle DM. }
	\label{tab:summary}

\begin{tabular}{|l|l|l|l|l|} 
\hline
\textbf{DM candidate} & \textbf{signature} & \begin{tabular}[c]{@{}l@{}}\textbf{observation}\\\textbf{type}\end{tabular} & \begin{tabular}[c]{@{}l@{}}\textbf{observation}\\\textbf{targets}\end{tabular} & \begin{tabular}[c]{@{}l@{}}\textbf{key SKAO}\\\textbf{requirements}\end{tabular}  \\ 
\hline
WIMP                  & synchrotron        & continuum                                                                   & \begin{tabular}[c]{@{}l@{}}Milky-Way satellites,\\galaxy clusters, GC\end{tabular}    & \begin{tabular}[c]{@{}l@{}}short baselines\end{tabular}        \\ 
\hline
ALP                   & birefringence      & \begin{tabular}[c]{@{}l@{}}continuum/\\spectro-polarimetry\end{tabular}             & \begin{tabular}[c]{@{}l@{}}quasars, AGN,\\strong lensing\end{tabular}                                                                   & \begin{tabular}[c]{@{}l@{}}polarization,\\long baselines,\\monitoring\end{tabular}             \\ 
\hline
ALP                   & \begin{tabular}[c]{@{}l@{}}stimulated \\ decay \end{tabular}              & spectral line                                                               & \begin{tabular}[c]{@{}l@{}}see WIMPs, + echoes\\ of bright sources\end{tabular}          & \begin{tabular}[c]{@{}l@{}} spectral resol., \\ short baselines \end{tabular}                                                             \\ 
\hline
\begin{tabular}[c]{@{}l@{}}ALP\\Dark photon \end{tabular}      & conversion         &               \begin{tabular}[c]{@{}l@{}}spectral line \\ continuum/transient  \end{tabular}                                               &\begin{tabular}[c]{@{}l@{}} neutron stars, Sun  \\   pulsars     \end{tabular}                                                   & \begin{tabular}[c]{@{}l@{}} spectral resol.     \\ time resol.\end{tabular}                                                          \\
\hline
\end{tabular}
\end{table}

\section{WIMPs}\label{sec:WIMP}

WIMP particles are weakly coupled to ordinary matter. In the standard WIMP scenario, they are stable but may pair--annihilate into certain species of SM particles.
Indirect detection of WIMPs in DM halos focuses on the search for a WIMP-induced component in the local anti-matter cosmic-ray fluxes and for an excess in Galactic or extragalactic fluxes of photons and neutrinos (e.g., \cite{Cirelli:2024ssz} for a recent review).

For cold DM, the energy of the annihilation process and the resulting products is set by the DM mass, approximately in the GeV-TeV range for WIMP DM. 
The production of GeV-TeV $e^+/e^-$ from prompt or subsequent processes can lead to synchrotron radiation in the radio band when interacting with a $\mu$G magnetic field, as is typical for galaxies.

The source term associated with the production of $e^+/e^-$ is given by:
\begin{equation}
q^{\rm ann}_e(E,r)=\langle\sigma_{\rm ann} v\rangle\,\frac{\rho_{\rm DM}(r)^2}{2\,M_{\chi}^2} \times \frac{dN_e^{\rm ann}}{dE}(E) \;,
\label{eqQ}
\end{equation}
where $\langle \sigma_{\rm ann} v\rangle$ is the velocity-averaged annihilation rate, $\rho_{\rm DM}(r)$ is the DM halo mass density profile at the radius $r$ (we assume spherical symmetry), $M_{\chi}$ is the mass of the DM particle, and $dN_e^{\rm ann}/dE$ is the number of electrons/positrons emitted per annihilation in the energy interval $(E,E+dE)$, obtained by weighting spectra for single annihilation channels over the corresponding branching ratio.
\footnote{In the case of WIMPs as non self-conjugate particles (Dirac fermions, complex scalars), the overall factor becomes 1/4, while 1/2 holds for the more common cases of self-conjugate particles (Majorana fermions, real scalars, real vectors).}

The annihilating WIMP source scales with the number density of WIMP pairs locally in space, i.e., with $\rho_{\rm DM}^2/2\,M_{\chi}^2$. 
The DM density profile is reconstructed from kinematic data, possibly including 21 cm data which can be collected at the SKAO, see the HI Galaxy Science Chapters in this book.

Electrons and positrons are nearly monochromatic if they are direct products of the annihilation (since the annihilating particles are essentially at rest); otherwise they have much broader spectra.
If the two-body final state particles from WIMP annihilations are predominantly quarks or weak gauge bosons, the injection of $e^+/e^-$ is mainly associated to a chain of hadronization and decay processes, leading to the production of charged pions and their subsequent decays into muons and in turn into $e^+/e^-$.
When instead the process of annihilation dominantly produces leptons, then $e^+/e^-$ are mainly originated directly from decays and have harder spectra and larger branching ratios.

Another possibility is that DM particles have a long but finite lifetime, and electrons and positrons arise from DM decays.
In this case, the source function takes the form:
\begin{equation}
q_e^{\rm dec}(r,E)=\Gamma_{\rm dec}\,\frac{\rho_{\rm DM}(r)}{M_{\chi}} \times \frac{dN_e^{\rm dec}}{dE}(E) \;,
\label{eqQ2}
\end{equation}
where $\Gamma_{\rm dec}$ is the decay rate and $dN_e^{\rm dec}/dE$ is the number of $e^+/e^-$ emitted per decay in $(E,E+dE)$.

The goal of the radio observations discussed in this Chapter is to derive constraints in the plane $M_{\chi}$-$\langle \sigma_{\rm ann} v\rangle$ and $M_{\chi}$-$\Gamma_{\rm dec}$ for given DM particle models providing $dN_e/dE$ or for a few benchmark choices of $dN_e/dE$.

\subsection{Synchrotron emission}\label{sec:synchro}
The transport of high-energy electrons and positrons injected by DM in an astrophysical medium can be described as a diffusive process governed by the equation:
\begin{equation}
-\frac{1}{r^2}\frac{\partial}{\partial r}\left[r^2 D\frac{\partial f_e}{\partial r} \right] 
  +\frac{1}{p^2}\frac{\partial}{\partial p}(\dot p p^2 f_e)=
  s_e( r, p)
\label{eq:transp}
\end{equation}
where we assume spherical symmetry and stationarity. In Eq.~\ref{eq:transp}, $p$ is the modulus of the momentum and $f_e(r,p)$ is the equilibrium $e^++e^-$ distribution function, related to the number density $n_e$ by: $n_e(r,E)dE=4\pi \,p^2f_e(r,p)dp$. Analogously, for the DM source function $s_e$, we have $q_e(r,E)dE=4\pi \,p^2\,s_e(r,p)dp$. 
The first term on the left-hand side describes the spatial diffusion, with $D(r,p)$ being the diffusion coefficient. The second term accounts for the energy loss due to radiative processes. In Eq.~\ref{eq:transp}, we neglected diffusion in momentum space (see below for their effect in galaxy clusters) and convection, since their effects are often subdominant in the context of DM searches.

The total synchrotron emissivity at a given frequency $\nu$ is obtained by folding the $e^++e^-$ number density $n_e$ with the total radiative emission power $P_{synch}$ over the electron/position energy $E$ \citep{Rybicki}:
\begin{equation}
    j_{synch}(\nu,r)=\int dE\,P_{syn}(r,E,\nu)\, n_e(r,E)
\label{eq:jsynch}
\end{equation}
For the typical systems considered in WIMP searches, absorption along the line of sight and within the source can be disregarded.

The flux density measured with the SKA telescopes can be estimated as
\begin{equation}
S_{th}(\nu,\theta_0) =\int d \phi\,d\theta\,\sin\,\theta\,\mathcal{G}(\theta,\phi,\theta_0)\int d\ell\,\frac{j_{synch}(\nu,r(\ell,\theta,\phi))}{4\pi}\;,
\label{eq:Isynch}
\end{equation}
where $\ell$ labels the coordinate along the line of sight, $\theta$ and $\phi$ describe the polar and azimuthal angles with respect to the pointing, $\theta_0$ is the direction of observation, i.e. the angular off--set with respect to the DM distribution center, and $\mathcal{G}$ provides the shape of the SKA synthesized beam.
We compare the theoretical prediction of Eq.~\ref{eq:Isynch} to foreseen observations with the SKA telescopes to derive constraints on the annihilation (decay) rate $\langle\sigma_{\rm ann} v\rangle$ ($\Gamma_{\rm dec}$) introduced in Eq.~\ref{eqQ} (Eq.~\ref{eqQ2}) at different $M_\chi$.

A crucial ingredient to be able to compute the synchrotron emissivity in Eq.~\ref{eq:jsynch} is the magnetic field. Its strength and profile can be investigated using polarization data. The studies described in the Cosmic Magnetism Chapters of this book offer an important synergy to the search for WIMPs.

\subsection{Targets}
A wide range of different cosmic structures can be high-density reservoirs of DM particles, and therefore possible targets for searching for DM annihilation signals.  \cite{Colafrancesco:2015ola} studied a large variety of cosmic structures at different redshifts, from dwarf galaxies to galaxy clusters, and investigated the observational prospects of the SKAO. Here we review the most promising structures and focus on the specific properties of the SKA telescopes in the AA* and AA4 configurations, building upon results of recent studies.

\subsubsection{Galaxy clusters}
Galaxy clusters are the largest virialized structures in the Universe, and therefore can host DM halos with the highest mass. This fact makes clusters promising candidate targets for detecting signals originating from DM annihilation \citep{2006A&A...455...21C}. Once the electrons produced in DM annihilation interact with the intra-cluster magnetic field, they can produce diffuse radio emission with properties resembling those observed in several tens of galaxy clusters, known as radio halos \citep{2012A&ARv..20...54F}. 
The spectral and spatial shape of radio halos can constrain WIMP DM properties (e.g., \citealp{2011A&A...527A..80C}).

However, in galaxy clusters other processes can accelerate relativistic electrons, such as hadronic interactions between cosmic ray protons and the nuclei of the thermal gas \citep{1999APh....12..169B} and shocks and turbulence following major cluster merging events \citep{1993MNRAS.263...31T,2001MNRAS.320..365B}. In particular, radio halos are usually observed in disturbed clusters, therefore diffuse stochastic acceleration related to turbulence may have a major role in producing the radio halos (e.g., \citealp{2013ApJ...777..141C}). While in principle electrons produced in DM annihilation can create radio halos for the observed properties of the intra cluster magnetic field and without violating gamma-ray upper limits from the \textit{Fermi}-LAT telescope~\citep{2016JCAP...11..033M}, distinguishing this emission from astrophysical processes is a difficult task.

Moreover, electrons produced in DM annihilation can be the seed of diffuse acceleration related to turbulence \citep{2019MNRAS.488.1401M}. This process can modify the shape of the electrons spectrum depending on the intensity and the duration of the acceleration phase, making it difficult to reconstruct the spectral profile of the seed electrons, and therefore the properties of the DM particles, from the spectrum of the observed radio emission.

There are two strategies to disentangle these astrophysical effects:
\textit{i)} observing the diffuse radio emission in relaxed clusters, where the level of turbulence is expected to be low, and the spectrum of the electrons should closely resemble that of seed electrons; \textit{ii)} observing the diffuse emission at high frequencies ($\nu\geq5$ GHz), where the energy losses are expected to dominate on turbulent acceleration even in disturbed clusters \citep{Marchegiani2025}, and the observed spectrum should be linkable to seed electrons.
Both strategies are challenging because the level of the WIMP-induced diffuse emission in relaxed clusters and at high frequencies is expected to be low. In the following, we provide some estimates about the conditions for which the SKAO in the AA4 baseline design can detect this emission.

We consider a galaxy cluster with the properties of the Coma cluster, but located at $z=0.2$. At lower redshifts, the angular extension of such a cluster might be challenging for an interferometer because of the lack of the zero-spacing in the dishes configuration and the consequent lack of sensitivity on large scales, especially at high frequencies \citep{2015aska.confE..75F}. For this illustrative cluster, we calculate the expected radio emission 
using the equations outlined in Sect.~\ref{sec:synchro}, but including the effect of turbulent acceleration \citep{2019MNRAS.488.1401M}.
We adopt a magnetic field as resulting from the Rotation Measures (RM) in the Coma cluster \citep{2010A&A...513A..30B}, and assume benchmark WIMP with mass 125 GeV and annihilation channel $b \bar b$ annihilating at a rate corresponding to the upper limit obtained by \textit{Fermi}-LAT in dwarf galaxies \citep{2015PhRvL.115w1301A}. We also assume that the emission expected from the main halo is boosted because of the presence of smaller subhalos, parametrizing this effect with a multiplicative boosting factor ${\cal B}$, which we fix at a value of ${\cal B}=70$ \citep{2014ApJ...788...27I}. The effect of turbulent acceleration is calculated by assuming a diffusion coefficient in the momentum space $D_{pp}=\chi p^2/2$ with a typical value of $\chi=5\times10^{-17}$ s$^{-1}$, and letting the electron spectrum evolve for various duration times $T_{acc}$ of the order of several $10^8$ yrs, as expected for a turbulent acceleration phase following a major merging \citep{2007MNRAS.378..245B}.

The forecasts are computed using the SKA sensitivity calculator \footnote{\href{https://sensitivity-calculator.skao.int/}{https://sensitivity-calculator.skao.int/}} for the SKA-Low telescope and for the bands 1, 2, and 5a of the SKA-Mid telescope in the AA4 configuration. 
Since the emission from DM annihilation is expected to be extended and faint, we consider configurations allowing a relatively large size of the synthesized beam. At $z=0.2$, the typical radius for a radio halo of $R_H=500$ kpc corresponds to $\theta_H\sim146$ arcsec: we select configurations providing a synthesized beam in the range 10-40 arcsec, so that the cluster size can be sampled with a sufficient number of beams. 
For SKA-Low we use Briggs weighting with robustness parameter 1, while for SKA-Mid we use uniform weighting with tapering of 32, 16.5, and 13.7 arcsec for bands 1, 2, and 5a, respectively. With such beam sizes, the instrument noise is expected to be dominated by the confusion noise, so we adopted a reference value of 1 hour of exposure time, verifying that the noise is not expected to significantly decrease even for much longer exposure times. The tapering minimizes the importance of the longest baselines such that the sensitivity limits predicted for AA* are not very different from those in the AA4 configuration. For  bands 1 and 2 of SKA-Mid we adopt the configurations with only the 15-m antennas in order to maximize the frequency extension of the band. The sensitivity limits on the flux density are estimated through (see \citealp{2012A&A...548A.100C}):
\begin{equation}
f_{min}\sim 1.2\times10^{-4} \xi_1 \left(\frac{F_{rms}}{10\,\mu Jy}\right)\left(\frac{100\,\mbox{arcsec}^2}{\theta_b^2} \right) \left(\frac{\theta_H^2}{\mbox{arcsec}^2} \right)\;,
\label{eq:fluxlimit}
\end{equation}
where the value of $F_{rms}$ and the beam size $\theta_b$ are provided by the sensitivity calculator, and we assume $\theta_H=146$ arcsec and a flux detection limit given by twice the RMS, i.e., $\xi_1=2$.

\begin{figure}
\vspace{-1.cm}
    \centering
 \includegraphics[width=0.7\columnwidth]
 {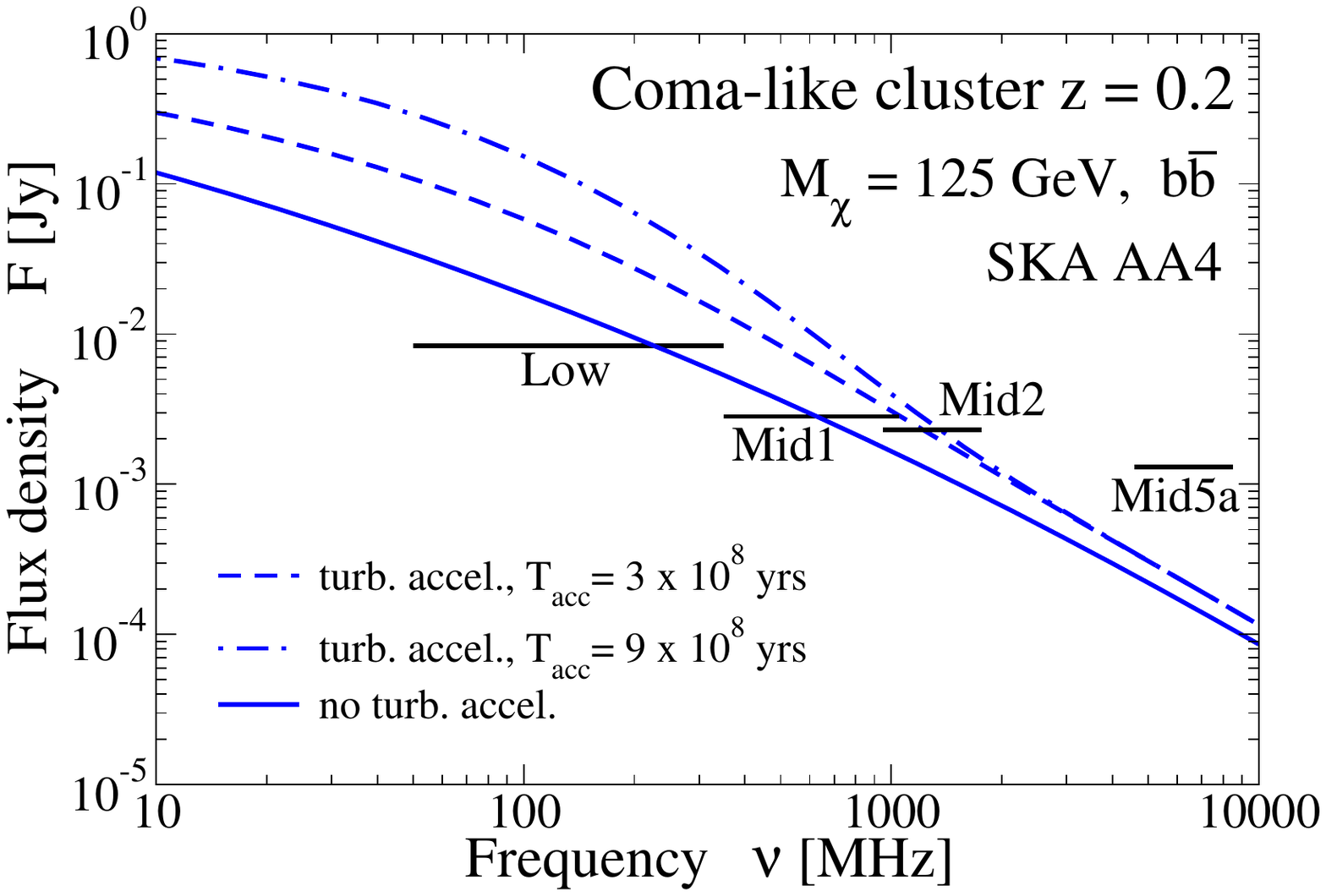}
    \caption{Spectrum of the diffuse radio emission calculated for a Coma-like cluster located at $z=0.2$ from the annihilation of a WIMP particle with mass 125 GeV and annihilation channel $b {\bar b}$, for a cross section of $\langle \sigma_{\rm ann} v\rangle=3\times10^{-26}$ cm$^3$ s$^{-1}$ and boost factor ${\cal B}=70$, without turbulent acceleration (solid blue line), and with turbulent acceleration with intensity $\chi=5\times10^{-17}$ s$^{-1}$ and duration $T_{acc}=3\times10^8$ yrs (dashed blue line) and $T_{acc}=9\times10^8$ yrs (dot-dashed blue line). The 95\% C.L. limits for the SKA-Low telescope and SKA-Mid telescope in 1, 2, and 5a bands (horizontal black lines) are calculated for 1hr of observations using the SKA calculator in the AA4 baseline design as described in the text.}
    \label{fig:clusterz02}
\end{figure}

In Fig.~\ref{fig:clusterz02} we plot the expected radio spectra without and with turbulent acceleration for durations of the acceleration phase of $3\times10^8$ and $9\times10^8$ yrs, compared with the estimated sensitivities. The 5a band of SKA-Mid is where the derivation of DM properties could be more accurate, since the effect of turbulent acceleration is less important, but also where the sensitivity looks not sufficient.
%
At lower frequencies the possibility of detecting the WIMP-induced emission is higher, especially in presence of turbulent acceleration. However, the spectral shape is modified depending on the intensity and the duration of the acceleration phase. 
The solid line in Fig.~\ref{fig:clusterz02} also represents a DM forecast for relaxed clusters. 
%


The expected synchrotron emission arising from WIMP annihilation in galaxy clusters is well within the reach of the SKAO AA4 and AA*.
To distinguish DM emission from the one coming from
other astrophysical processes, a good characterization of the diffuse emission spatial distribution is necessary \citep{2015MNRAS.452.1328M}.
Also, suppressing systematic uncertainties related to astrophysical and DM model predictions will be crucial to set strong limits on WIMPs. Astrophysical uncertainties can be minimized in targets where the magnetic field, cosmic-rays, and DM density will be well traced, as foreseen by other Chapters in this book. Similarly, stacking several clusters will help to disentangle the nearly universal properties of the DM signal.

\subsubsection{Milky Way satellites}
Dwarf galaxies are a prime target for WIMP DM searches. This is due to their DM content, lack of strong baryonic backgrounds, and relative proximity. The disadvantages are the lack of magnetic field information for such objects and that their small size renders them diffusion dominated. The latter also increases the computational costs of solving the diffusion-loss equation. Several searches have already been conducted on such objects with a variety of radio telescopes~\citep{Natarajan:2013,Spekkens:2013,Regis:2014tga,Natarajan:2015,Regis_2017,Kar_2019,Cook:2020,Vollmann:2020a,Basu_2021,Gajovic:2023bsu,Guo_2023}. However, extant dwarf galaxy searches have so far struggled to produce strongly competitive limits due to the uncertainty on the magnetic properties that leads to an uncertainty of several orders of magnitude in the predicted radio signal in dSphs~\citep{Regis:2014tga}, and thus in the bounds on the WIMP annihilation rate.
SKAO can change this picture.

To derive SKAO capability, we pick a dwarf galaxy that was previously observed to have no diffuse emission with ATCA~\citep{Regis_2017}, namely Reticulum II, and produce non-detection constraints according to the nominal SKA sensitivity in both AA4 and AA* configurations. 
Although limits from radio observations tend to be more competitive for WIMPs annihilating in leptonic channels, we use the b-quark channel since it is more typical for supersymmetric models and it is often used as standard point of comparison across all targets and frequencies. The Reticulum II galaxy is a very suitable target given its combination of proximity and DM abundance, as well as a lack of diffuse emission. 

To produce accurate projections we simulate a noise map (assuming no observed diffuse emission) via a normal distribution of fluxes with an RMS matching SKA sensitivity and an average $\ll$ the RMS. We then convolve this map with a chosen beam size of 1', matching the ATCA data for purposes of comparison. This noise map is truncated at $\pm 3$ times its own RMS to prevent outliers dominating the results.  We then produce a WIMP prediction map, at the same resolution, via the DarkMatters tool~\citep{2025PDU....4701745S} to obtain a 95\% confidence interval limit on $\langle\sigma_{\rm ann} v\rangle$. We use predictions and noise that are band-averaged in three ranges: 50 to 300 MHz (SKA-Low), with 10 hour sensitivity $8.6$($22$) $\mu$Jy/beam for AA4(AA*); 350 to 900 MHz (SKA-Mid Band 1), similar to the MeerKAT UHF band, with $2.5$ $\mu$Jy/beam; and 1.1 to 3.1 GHz (SKA-Mid Bands 2\&3), similar to ATCA observations in \cite{Regis_2017}, with $0.51$ $\mu$Jy/beam. To ensure statistical robustness, we perform the procedure for a large number of random noise realisations and average their resultant limits.  We validate the procedure against the results of \cite{Regis_2017} and find that an RMS of $\approx 15$ $\mu$Jy/beam produces predictions in good agreement. This suggests that this prediction model is relatively robust, as the sensitivity required is similar to the RMS of the initial ATCA maps.

\begin{figure}
\vspace{-1.cm}
    \centering
 \includegraphics[width=0.7\columnwidth]
    {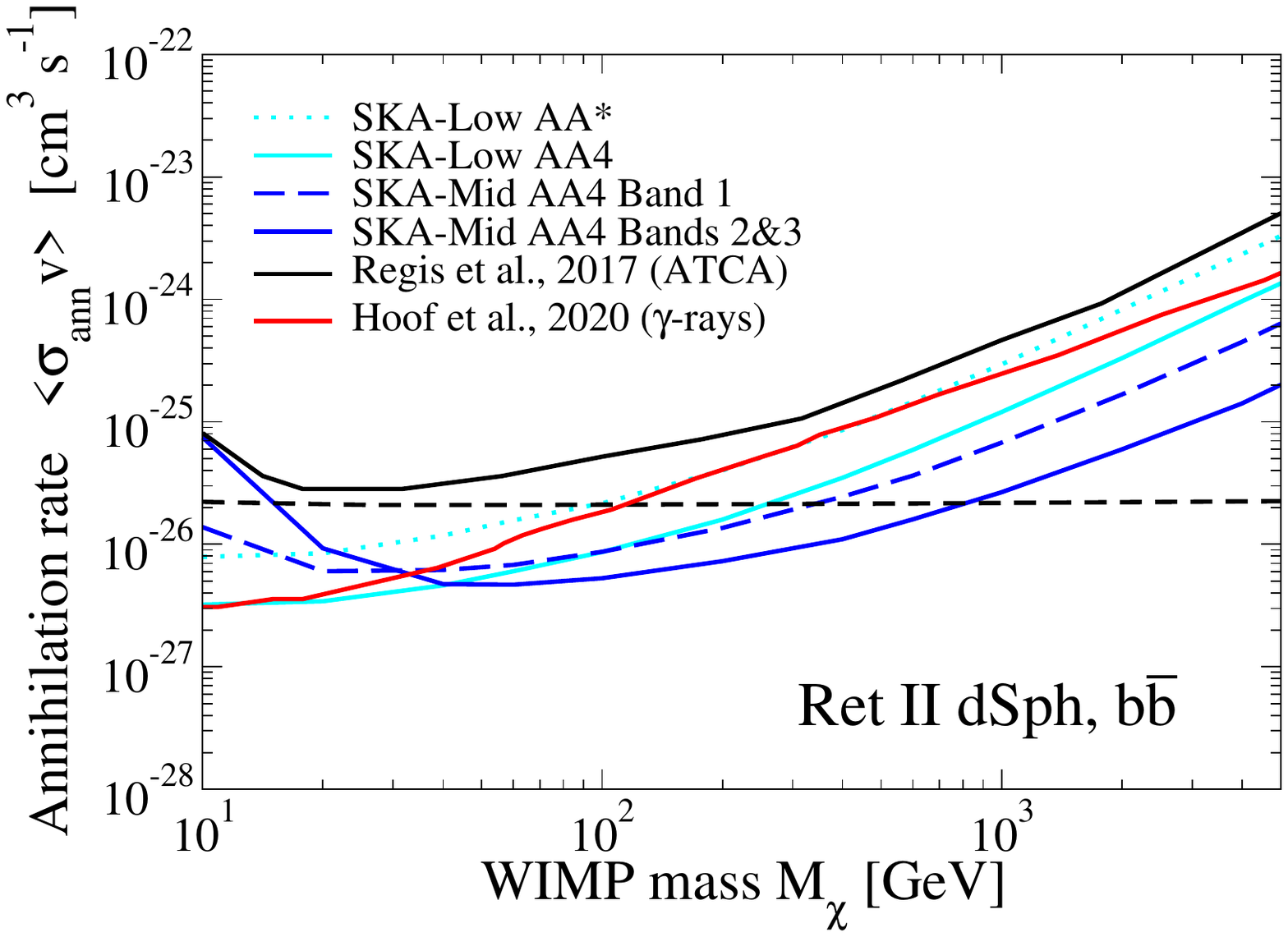}
    \caption{Projected non-detection limits on the $b\bar{b}$ channel for 10 hours of observing Reticulum II. The red curve represents Fermi-LAT limit from a combined set of dwarf galaxies~\citep{Hoof_2020}, the black curve is from radio observations with ATCA~\citep{Regis_2017} and the black dashed line is the thermal relic value.}
    \label{fig:ret2-ska}
\end{figure}

Using the fiducial model from \cite{Regis_2017} we find limits displayed in Figure~\ref{fig:ret2-ska}. We see that SKA AA$^*$ design improves upon ATCA limits by more than an order of magnitude when multiple bands are considered. Thus, the frequency range of the SKA telescopes is a major contributor in allowing for far more constraining results over a wide range of WIMP masses. 
For the given band-averaging, these results do not differ substantially between AA4 and AA$^*$ in the SKA-Mid bands. However, SKA-low limits are a factor of $\sim 3$ worse for AA$^*$ than AA4.   
Importantly, the maximum mass where the thermal cross section is excluded is close to 1 TeV, nearly an order of magnitude better than the combined gamma-ray analysis of a set of dwarf galaxies. Thus, the projected SKA performance allows for cutting edge limits on WIMP properties even in the intermediate AA* configuration. 
This suggests that valuable DM searches can be conducted in dwarf galaxies with both AA4 and AA* configurations, as only the low frequency results are weakened for AA*. 

Dwarf galaxies are not the only satellite systems of the Milky Way. Globular clusters are thought to have only a sub-dominant DM component, but if the halo has a dense DM distribution they can provide a strong annihilation signal. Some globular clusters, such as Omega Centauri, are plausibly explained as the core of a tidally disrupted dwarf galaxy (e.g. \citealp{oc-dsph-4}) supporting the possibility of non-negligible DM content~\citep{Reynoso_Cordova_2022}. This would result in very strong constraints in both radio and gamma-rays~\citep{Beck:2023ZY}. 

The largest Milky Way satellites are the Magellanic Clouds. They are DM dominated systems and have a significant magnetic field, therefore the synchrotron emission induced by WIMPs is expected to be bright. Competitive bounds on WIMPs have been obtained with ASKAP data in \cite{Regis_2021}.
Observations with the SKAO are expected to probe the thermal cross-section level (dashed line in Fig.~\ref{fig:ret2-ska}) in the multi-TeV regime. The disadvantage with respect to dwarf spheroidals is that detailed projected constraints are not straightforward to compute as they depend on the sensitivity the SKA telescopes to diffuse emission and on the LMC astrophysical emission.

Dark subhalos (i.e., DM structures with no luminous counterpart) are expected to be copiously present in our Galaxy, according to numerical simulations with CDM. On the other hand, uncertainties on their location and on magnetic properties prevent a reliable determination of the associated projected constraints. Despite the difficulty in setting bounds, dark subhalos might provide an interesting channel for detection~\citep{Leite:2016}, and unidentified extended sources detected with the SKAO should be carefully scrutinized in this respect.

\subsubsection{Other targets}
Other promising targets include nearby massive galaxies, in particular if their magnetic field is significant and extended. M31, which is part of the Local Group, has been shown to provide interesting radio bounds~\citep{Egorov:2022fkc}.
The brightest WIMP-induced synchrotron source is expected to be our Galactic Center (GC)~\citep{Regis:2008ij}, however this requires a complicated assessment of various astrophysical backgrounds and systematic uncertainties.
Finally, the collective emission from all DM halos in the Universe provides an extragalactic, nearly-isotropic flux that can be investigated using data from SKAO surveys of large patches of the sky, by analyzing its small-scale anisotropies~\citep{Singal:2022jaf}. 

\subsection{Synergies with other indirect searches }

 There are a wide range of indirect detection searches looking for SM particles produced by DM decay or annihilation. As mentioned above, the immediate products might be not stable, and quickly decay or hadronise to stable particles: photons, neutrinos, electrons and positrons, protons and anti-protons, and heavier nuclei. With the SKAO we can search for evidence of DM annihilation via the synchrotron radiation of electron/positron annihilation products, which can set complimentary limits on DM models when compared or combined with other searches which are subject to different systematics. For a review of WIMP indirect searches, see, e.g., \cite{Cirelli:2024ssz}.

\begin{figure}[t]
\vspace{-1.cm}
    \centering
 \includegraphics[width=0.7\columnwidth]{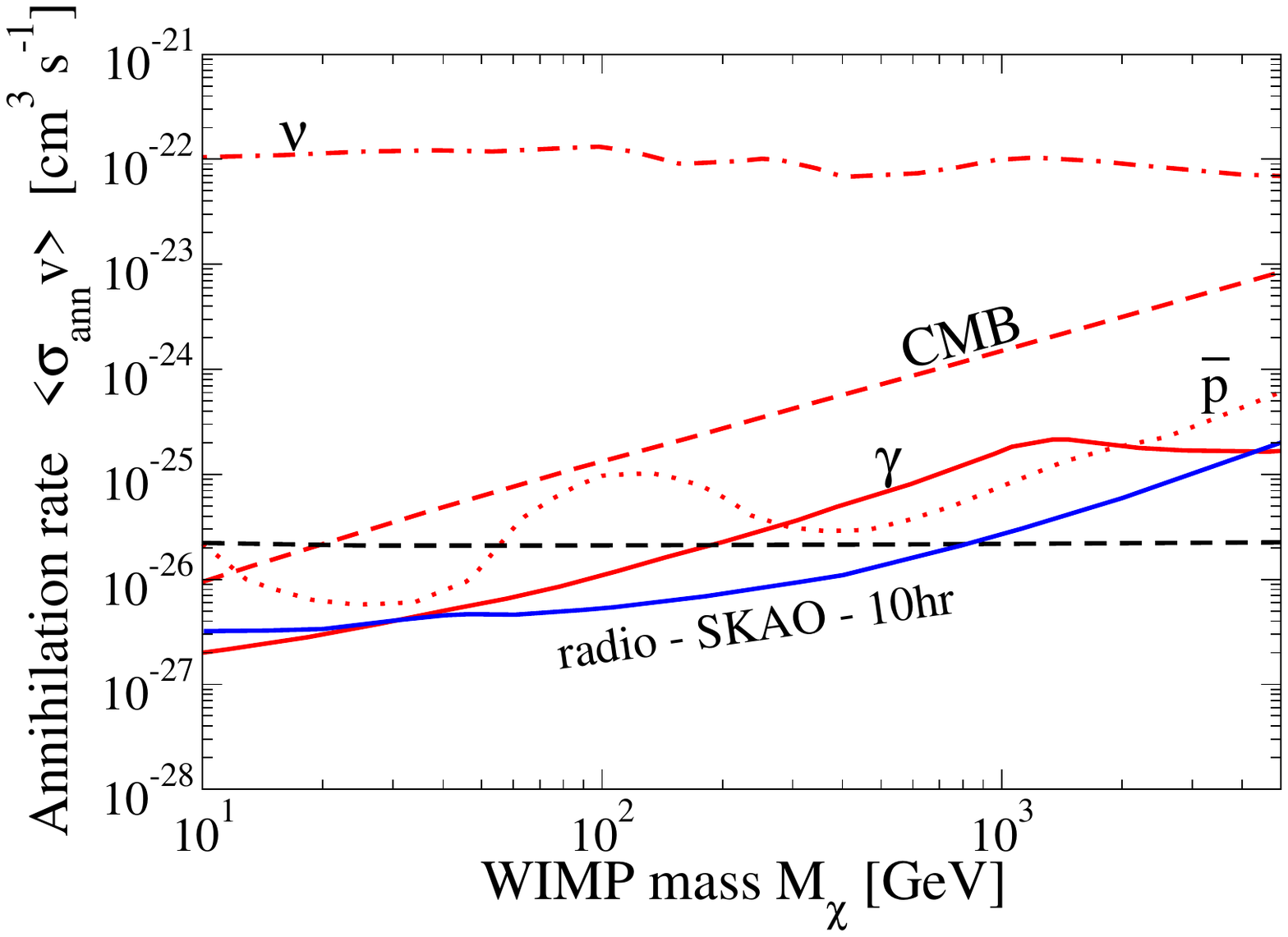}
    \caption{Comparison of WIMP limits obtained with different indirect detection techniques (gamma-rays - solid red; anti-protons - dotted red; CMB - dashed red; neutrinos - dashed-dotted red) and with the SKA telescopes in the AA4 baseline design (solid blue). We considered a WIMP annihilating into $b\bar b$, the compilation of indirect bounds in Fig. 6.14 in \cite{Cirelli:2024ssz} and 10 hours of SKA observation of a dwarf galaxy (see Fig.~\ref{fig:ret2-ska}).   
    }
    \label{fig:summaryWIMP}
\end{figure}

For example, DM annihilation may produce gamma rays via prompt or radiative emission.
Gamma ray telescopes such as Fermi-LAT~\citep{fermilat}, 
and the future CTA~\citep{cta} aim to search for two types of spectra: (1) continuum spectra, and (2) gamma ray lines produced by the direct annihilation of DM into photons.
Bounds are comparable to existing radio bounds, but often considered more robust since they do not depend on the description of the magnetic properties of the system.

Antimatter searches also provide stringent bounds on WIMPs. On the other hand, since charged particles diffuse in the magnetic field of our Galaxy, this randomizes their directions, and alters their energy spectrum, making more difficult to track back to the DM source and thus having larger systematic uncertainties than gamma-ray searches.

The injection of secondary particles produced by DM annihilation at $100 < z < 1000$  can produce heating of the medium and ionization of the hydrogen gas. This can affect the cosmic microwave background anisotropies, which strongly constrain light DM particles, and might be visible in measurements of the epoch of reionization with the SKAO.

A comparison between the SKA projected sensitivity and existing bounds from other indirect detection searches is reported in Fig.~\ref{fig:summaryWIMP} in the example of a WIMP annihilating into $b\bar b$.

Other detection strategies for WIMPs include scattering off targets in underground laboratories and production at particle colliders, e.g., \cite{Cirelli:2024ssz}. They are complementary to radio and other indirect searches, since related to different particle processes, which means also that the model constraining power is less straightforward to compare.

\section{Light bosonic DM }
As already mentioned in the Introduction, the radio signals associated to ALPs arise because of the presence of the  term $ -\frac{g_{a\gamma}}{4} a F_{\mu \nu}\tilde{F}^{\mu \nu}$ 
in the Lagrangian.
This coupling involves three fields: the ALP field $a$, the electromagnetic field $F_{\mu \nu}$ and its dual $\tilde{F}^{\mu \nu}$. It gives rise to the decay of ALPs into two photons, conversion of an ALP into a photon in the presence of an external magnetic field (or electric field, which is however typically subdominant in astrophysical environments), and the rotation of the polarization angle of a photon when interacting with an ALP. 

For what concerns the dark photon, the kinetic mixing $ -\frac{\epsilon}{2} F_{\mu \nu}F^{\mu \nu}_{DP}$ between the electromagnetic field and the dark photon field $F^{\mu \nu}_{DP}$, leads to a possible conversion  of a dark photon into a visible photon. The signature is similar to the one of ALP conversion and will be discussed in Section~\ref{sec:ALPcon}.

\subsection{Decay} \label{sec:ALPdec}

The decay rate of ALPs into photons in vacuum is proportional to $g_{a\gamma}^2\,m_a^3$.
At radio frequencies, i.e., $m_a\lesssim 10^{-5}$ eV, and for allowed values of $g_{a\gamma}$,
this decay rate is too small to lead to any detectable signal. However, the decay rate is enhanced in the presence of a background of photons with energy equal to half the ALP mass~\citep{Arza:2019nta, Caputo:2018vmy}. This is due to Bose enhancement, as the final state of the process is highly occupied. This process is commonly referred to as stimulated decay, in analogy to stimulated emission.

Sources of stimulating radiation can be divided in isotropic, from the point of view of the observer, and directional.
The isotropic radio flux is mainly provided by the cosmic microwave background and the extragalactic radio background~\citep{2011ApJ...734....5F}, while directional sources include diffuse synchrotron radiation in our Galaxy, the radio emission from supernova remnants (SNRs), and radio galaxies.

Let's consider a directional source emitting a background of stimulating photons with momentum $\vec{k}$, and an ALP with momentum $\vec{p}$ in the observer's rest frame. One of the photons produced in the decay will be identical to the stimulating one. In particular, it will carry the same momentum $\vec{k}$. We will call the ensemble of such photons ``collinear emission". By conservation of momentum, the other photon produced in the decay will have momentum $-\vec{k}+\vec{p}$, and travel nearly in the opposite direction as the stimulating one. It will contribute to the ``echo" (or gegenschein) of the stimulating source. 
Therefore, three signatures can be identified~\citep{Sun:2023gic, Todarello:2023xuf}: one collinear emission, and two echos. The first kind of echo, the backlight echo, arrives to the observer from the direction opposite to the source. The second kind of echo, the frontlight echo, comes from behind the source.
 
Importantly, the direction of propagation of the echos is not exactly opposite to that of the stimulating radiation, but is modified by the DM momentum. For example, a DM velocity dispersion perpendicular to the line of sight $\sigma_\perp$ implies that the signal is spread on the sky over an angular extent $\delta\theta \sim 2\sigma_\perp$~\citep{Sun:2021oqp, Buen-Abad:2021qvj}. For ALP decays in the Milky Way, the echo signal is smoothed over $ \sim 10'$, leading to a reduced sensitivity, especially for interferometer observations, compared to the idealized case of no DM velocity dispersion.

Imposing conservation of energy, for non-relativistic ALPs, we obtain that the energies of the emitted photons are $(m_a \pm p_\parallel)/2$, where $p_\parallel$ is the component of $\vec{p}$ along the direction of $\vec{k}$. As a consequence, the signal has a relative spectral width  of order the DM velocity dispersion along the line of sight $\sigma_\parallel$. In the following, we will consider ALP DM contained in the dark halos of our Galaxy, dwarf galaxies, and elliptical galxies. The velocity dispersion $\sigma$ ranges from $10^{-5}c-10^{-4}c$ for dwarfs to $10^{-3}c$ for galaxies. Galaxy clusters are also a possible target.

Under certain conditions, the intensity of radiation emitted by ALP stimulated decay can grow exponentially due to parametric resonance~\citep{Tkachev:1986tr}. The growth rate is $s \sim g_{a\gamma}\sqrt{\rho_a} / (2\sqrt{2})$, where $\rho_a$ is the ALP energy density. This mechanism is ineffective for the smooth component of the DM in the Milky, as to gravitational redshift ``de-tunes" the resonance~\citep{Arza:2020eik}. However axion stars, gravitationally bound clumps of ALPs condensed in  the ground state, may enable such growth~\citep{Hertzberg:2018zte, Levkov:2020txo} if $s R >1$, where $R$ is the size of the clump. Sensitivity forecasts for this signal with the SKA telescopes have been presented in~\cite{Maseizik:2024uln}. These projections rely on several model-dependent assumptions, particularly regarding the axion star mass function. We refer the reader to~\cite{Maseizik:2024uln} for further details.

\begin{figure}[h]
    \centering
	\includegraphics[width=0.5\columnwidth]{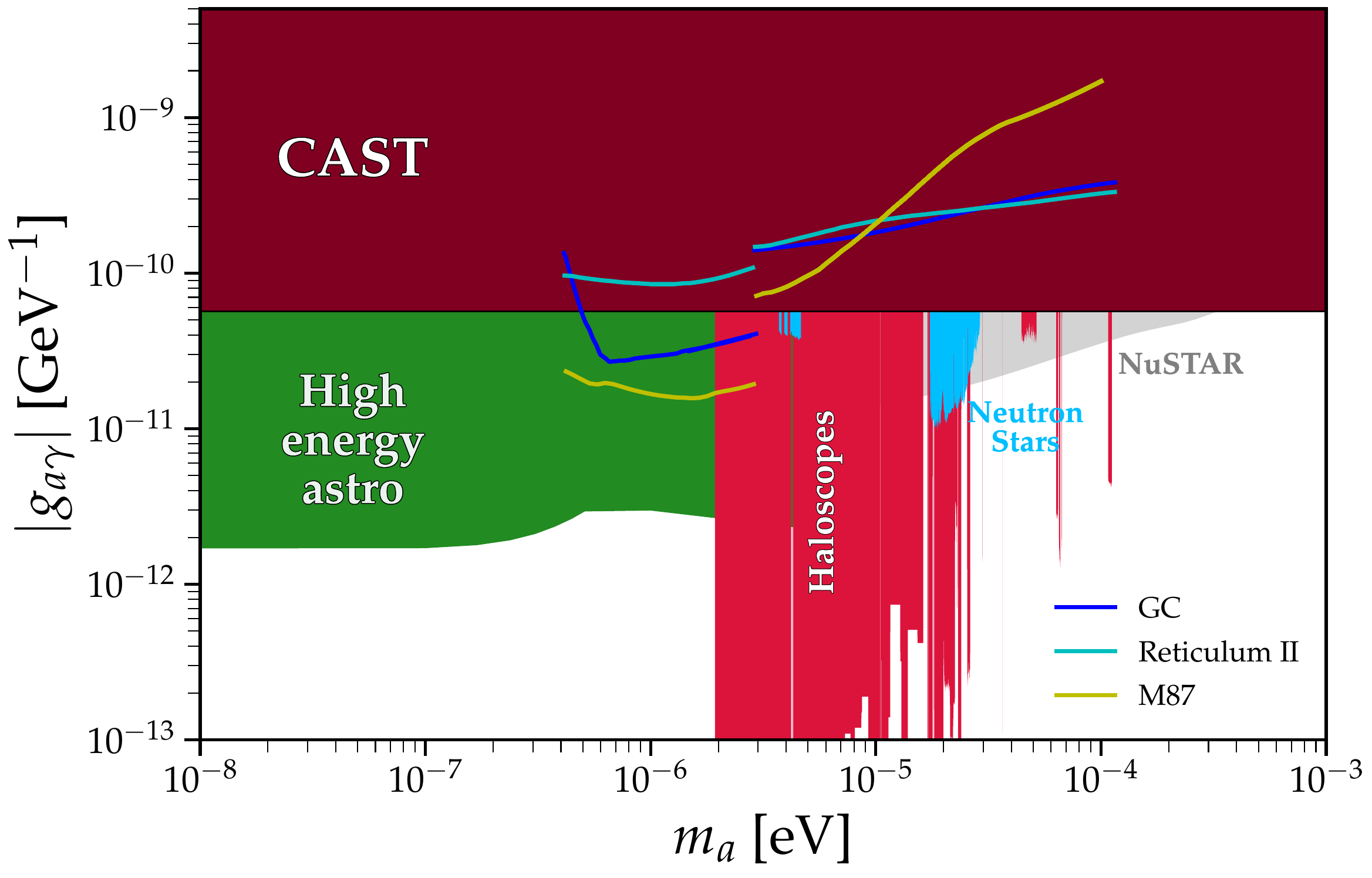}%
	\includegraphics[width=0.5\columnwidth]{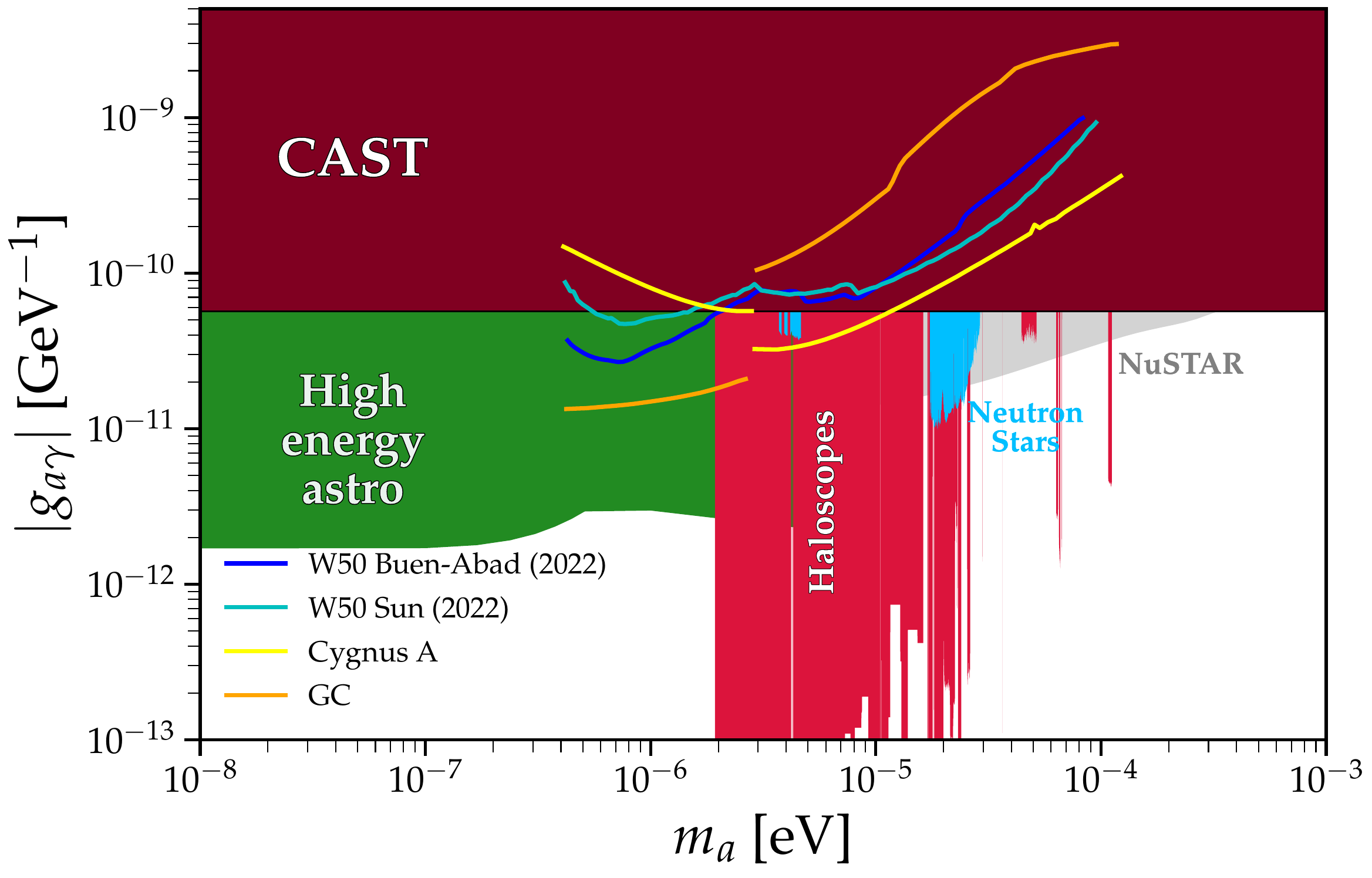}
    \caption{Collinear emission (left). Backlight echo (right). All have $t_{obs}=100$~hrs, signal-to-noise$=1$, AA4 configuration, and ALPs making up 100\% of the DM density.
    Forecasts for SKA observations are explained in the main text, while other limits are taken from the repository~\href{https://cajohare.github.io/AxionLimits/}{https://cajohare.github.io/AxionLimits/}.
    }
    \label{fig:ALPdecay}
\end{figure}

\textbf{Collinear emission} \cite{Caputo:2018vmy} consider the collinear emission from three DM-rich targets: the dwarf galaxy Reticulum II, M87, and the Galactic Center (GC). Sensitivity forecasts for the SKA telescopes in the AA4 baseline design are shown on the left panel of Figure~\ref{fig:ALPdecay}. 
In the case of Reticulum II, the sources of stimulating radiation are the CMB and the extragalactic radio background, while for the GC the dominant enhancement of the decay rate comes from the Galactic synchrotron background. These sources are rather diffuse and, at higher SKA-Mid frequencies, become broader than the largest angular scale that interferometer can image. The forecasts
are thus quoted for SKA-Mid operating in the single-dish mode. For SKA-Low, there's no such issue and the forecasts are for the array operating in the interferometric mode. 
In M87, the main stimulating radiation is the diffuse radio emission from the galaxy itself. This emission is mostly concentrated in the central region of the galaxy, and, in combination with the larger distance, it implies that this source is much more compact than Reticulum II and the GC. M87 forecasts for both SKA-Mid and SKA-Low are for the array operating in the interferometric mode.
While the collinear emission does not get spread across the sky due to the DM velocity dispersion like the echo, allowing more radiation to be collected with a single pointing of the telescope, any fore- or background continuum needs to be subtracted. The forecasts described above assume a perfect removal of the smooth continuum from the measurement.

\textbf{Backlight echo}
Existing sensitivity forecasts based on the backlight echo consider the stimulated decay of DM in our Galaxy. The explored stimulating sources are SNRs \citep{Buen-Abad:2021qvj, Sun:2021oqp}, the GC \citep{Dev:2023ijb} and the radio galaxy Cygnus A \citep{Ghosh:2020hgd}.
The angular broadening due to the DM velocity dispersion suppresses the signal on the longest interferometric baselines, leading to a reduction in overall sensitivity.
\cite{Sun:2021oqp} consider SKA telescopes in the interferometric mode only, while \cite{Buen-Abad:2021qvj} consider both single-dish and interferometric modes, reporting the most stringent constraint achievable for each ALP mass. The forecasts are shown on the right panel Figure~\ref{fig:ALPdecay} as blue and cyan lines. The primary source of uncertainty in these projections arises from modeling the time evolution of the stimulating source's luminosity. In fact, the echo signal depends on the entire history of the stimulating source. Echo photons reaching us today, originate from  decays stimulated by photons emitted by the source a time $t \sim (2d + x)/c$ ago, where $d$ is the distance to the decay location and $x$ the distance to the stimulating source. The total echo flux is the sum of the contributions from decay happened at all distances $d$ along the line of sight. SNRs are particularly promising as stimulating sources because their luminosity was significantly higher in the past.
The forecasts for Cygnus A is shown on the right panel of Fig.~\ref{fig:ALPdecay} as an orange line. \cite{Ghosh:2020hgd} consider the sensitivity of SKA2. Here we rescaled their projections by a factor of $\sqrt{10}$ to emulate AA4 sensitivity. The forecasts for the backlight echo from the GC using SKA telescopes in the autocorrelation mode are shown in orange on the same Figure.

In conclusion, ALP stimulated decay offers a robust way to explore the parameter space of the ALP-photon coupling. While ALP-photon conversion, discussed in the next subsection, might give stronger projected bounds on $g_{a\gamma}$, it also requires modeling the magnetic field and plasma in the environment being considered, as well as the DM phase space distribution. 
On the other hand, in the case of ALP decay, one needs to model only the DM distribution and the emission of the stimulating source, something that typically can be well characterized from observational data.

\subsection{Conversion} \label{sec:ALPcon}

One of the most powerful probes of ALPs is the search for photon signals arising from axion-photon conversion in strong magnetic fields. 
In the context of astrophysical searches, neutron stars (NSs) have been recognized as ideal targets~\citep{Pshirkov:2007st,Huang:2018lxq,Hook:2018iia}, 
sparking a lot of interest in recent years.
This relies on two crucial properties of NSs: their extremely large magnetic fields and the ambient plasma in the magnetosphere surrounding the neutron star, both of which enhance the conversion probability of ALP DM into photons.
In particular, the presence of a plasma enables resonant conversion in the regions where the ALP and photon dispersion relations become degenerate, i.e. where $m_a\simeq\omega_p,$ with $\omega_p$ the plasma frequency of the medium. 
Since the ALP DM particle being converted is non-relativistic, the resulting photon signal is a narrow line, centered at an energy around the ALP mass. Moreover, for the range of plasma frequencies around NSs, this emission falls within the radio band. As the plasma frequency decreases with distance from the neutron star, the search for this signal allows testing a range of ALP masses, each one associated with the conversion layer where the resonant condition is satisfied.
Another interesting phenomenon occurring in the neutron star magnetosphere is the production of ALPs from oscillations of the ambient electromagnetic fields. The resulting population of ALPs can lead to a variety of signals, including broadband radio fluxes and a narrow radio line.
Finally, the solar atmosphere constitutes another promising and complementary target to search for the conversion of light DM into radio photons, especially in the context of dark photon DM, as shown in~\cite{An:2020jmf}.
In the following, we discuss the experimental searches performed so far with current radio telescopes and the prospects for detection with the SKAO.
We will begin by discussing the case of the Sun, which, for our purposes, presents a simpler environment compared to NSs.

{\bf The Sun.} The probability for conversion of ALP DM into photons ($P_{a\gamma}$) in the solar atmosphere, and the resulting radio flux $S$ at Earth are given by:
\begin{equation}
P_{a\gamma} \simeq \frac{\pi}{2}\frac{g_{a\gamma}^2\,B_{\bot }^2}{v_a \, |\omega_p^{\prime}|}\,,~~~~~S=\int \frac{d\Omega}{4\pi\Delta\nu} \rho_a\, v_a\, P_{a\gamma}\, e^{-\tau}\,,
\label{eq:Sun}
\end{equation}
where $B_{\bot }$ is the component of the magnetic field transverse to the direction of propagation of the ALP, $\omega_p^{\prime}$ is the gradient 
along the ALP/photon trajectory of the plasma frequency, $\Delta\nu$ is the bandwidth, and $\rho_a$ and $v_a$ are the density and velocity of the ALP. All these quantities are computed at the layer of the solar atmosphere $r_c$ where the resonant condition is realized. 
To compute the signal flux, the angular integration is performed over the specific observations of the solar disk.
Absorption of solar radio photons during their propagation is accounted for by the optical depth $\tau,$ and two main processes are potentially important: inverse thermal bremsstrahlung and gyro-resonance absorption.
Finally, radio photons can experience significant refraction and scattering during their propagation in the solar atmosphere. At radio frequencies, this leads to an angular broadening of the emission of the order of 25-30\% for the whole Sun~\citep{Sharma:2020}.
The formalism described above can be easily applied to the case of another DM candidate, the dark photon. In this case, $P_{a\gamma}$ is obtained through the substitution~\citep{An:2023mvf}: $g_{a\gamma}\,B_{\bot } \rightarrow \sqrt{2/3}\,\epsilon\, m_{A^{\prime}},$ with $m_{A^{\prime}}$ the dark photon mass. As evident, in this case, an external magnetic field is not required in the conversion process. 

\begin{figure}[t]
    \centering
	\includegraphics[width=0.5\columnwidth]{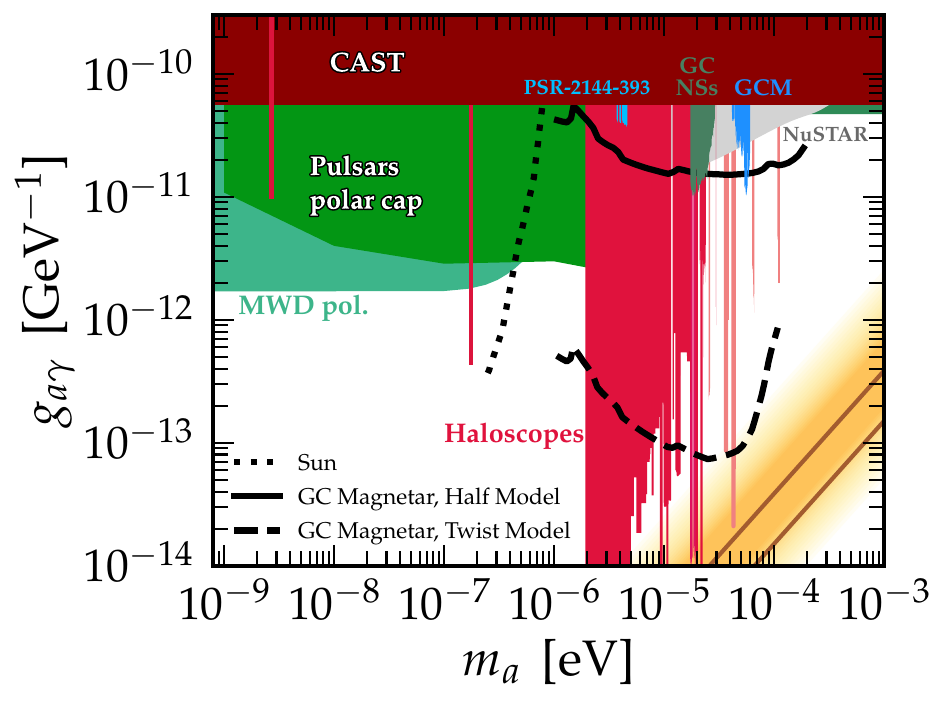}%
	\includegraphics[width=0.5\columnwidth]{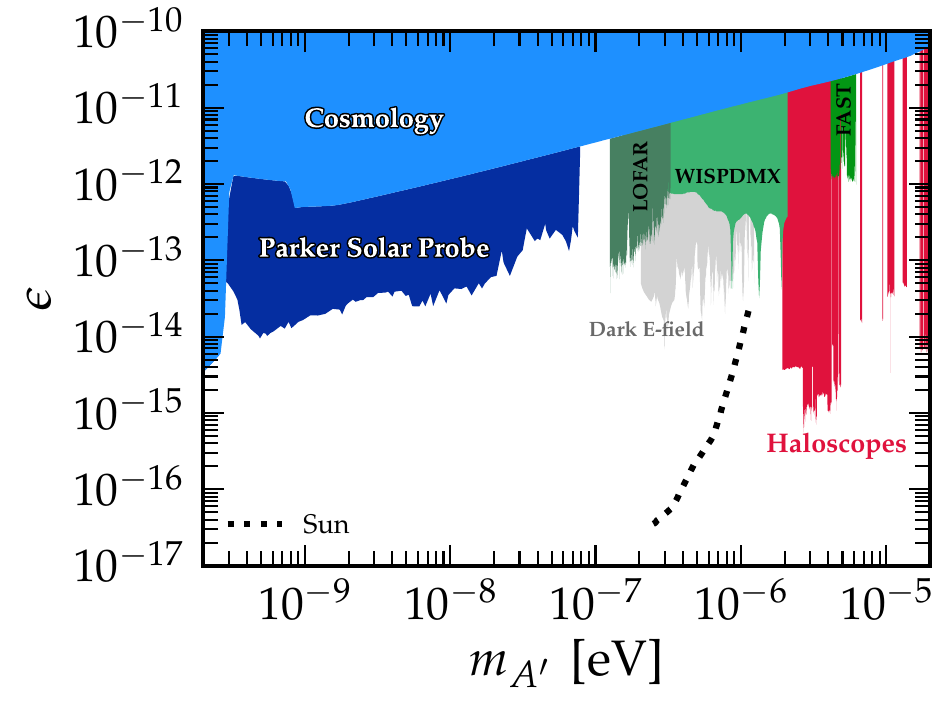}
    \caption{Existing bounds (colored areas) and projected sensitivities for the SKA telescopes in the AA4 configuration (black lines) on the axion-photon coupling (left panel) and dark photon kinetic mixing (right panel).
    Bounds from radio observations are explained in the main text, while other limits and the QCD axion band (yellow band in the left plot) are taken from the repository~\href{https://cajohare.github.io/AxionLimits/}{https://cajohare.github.io/AxionLimits/}.
    }
    \label{fig:Conversion}
\end{figure}

The conversion signal described above is a radio line with an intrinsic width of the order of $\mathcal{O}(v/c)^2\simeq 10^{-7}-10^{-6},$ where $v$ is the DM velocity dispersion. 
A search for such a signal has been performed using LOFAR solar observations in the frequency range 30-80 MHz~\citep{An:2023wij}, and data from the STEREO satellite and the Parker Solar Probe ($\simeq$ 3 kHz-20 MHz)~\citep{An:2024wmc}.
In the context of dark photon DM the resulting exclusion limits, presented in Fig.~\ref{fig:Conversion}, are particularly compelling, significantly improving over previous ones. Instead, for ALPs, the bounds from this search are weaker than existing constraints from laboratory experiments.
Prospects for detection with the SKAO have been studied in~\citep{Todarello:2023ptf,An:2020jmf,An:2023mvf}. Fig.~\ref{fig:Conversion} shows the projected sensitivity reach
of SKA-Low AA4 assuming 100 hours of observations of the solar disk, taken from~\cite{Todarello:2023ptf} and adapted to the case of dark photon as explained above.
As evident, solar observations with the SKAO have the potential to cover currently unexplored regions of the parameter space, especially in the case of dark photon DM.

{\bf Radio antennas.} Before discussing the case of NSs, we shall mention that the antennas of radio telescopes can serve as detectors for dark photon DM. In the antenna, ambient dark photons can be converted into photons, which are then collected by the receivers. \cite{An:2022hhb} has derived strong bounds using data from the FAST telescope, see Fig.~\ref{fig:Conversion}, and found significant prospects for improvement with the SKAO.

{\bf Neutron stars.} In the case of NSs, modeling the photon conversion signal requires a more complex approach than that described in Eq.~\ref{eq:Sun}. Recently, this topic has been the focus of intense research activity. A simple description of the neutron star magnetosphere is provided by
the Goldreich-Julian model, which gives a dipole magnetic field inclined with respect to the rotation axis of the neutron star, and the associated distribution of charges in the surrounding plasma. 
Refined calculations of the probability for resonant conversion $P_{a\gamma}$ in such highly-magnetized plasma have been presented in~\citep{McDonald:2023ohd,McDonald:2024uuh}, showing good agreement with numerical simulations~\citep{Gines:2024ekm}. The anisotropic plasma in the magnetosphere and the curved spacetime around the neutron star can strongly affect the propagation of photons from the conversion layer. These effects are important to determine the properties of the radio signal, namely its intensity as well as the angular, spectral, and temporal features. For example, the line broadening of the signal induced by the energy exchange with the plasma typically dominates over the one set by the DM velocity dispersion. To account for these effects and model the radiative transport in the magnetosphere, ray tracing methods have been employed, see~\cite{McDonald:2023shx} and references therein.

Searches of ALP DM conversion signals have been performed both from isolated stars~\citep{Foster:2020pgt,Darling:2020plz,Darling:2020uyo,Battye:2021yue,Zhou:2022yxp,Battye:2023oac} and from populations of NSs~\citep{Foster:2020pgt,Foster:2022fxn}.
Concerning the former targets,
a promising object is the Galactic Center magnetar (GCM). Being located at only $\simeq 0.2$ pc
from the Galactic center, the ambient DM density around the GCM could be orders of magnitude larger than the local one (a factor $\mathcal{O}(10^5)$ assuming a NFW density profile), although this enhancement is very uncertain. Moreover, the GCM magnetic field is very large, estimated to be $\sim 10^{14}$ G at the surface of the star (assuming a dipolar model). The most recent constraints from the GCM on the ALP conversion signal have been obtained in~\cite{Battye:2021yue} using VLA observations. We show them in Fig.~\ref{fig:Conversion} (for Model A of that reference). More recently, a refined calculation of the signal has been presented 
in~\cite{Roy:2025mqw}, which has derived projected sensitivity for the SKAO. In Fig.~\ref{fig:Conversion} we show their results for two different models of the magnetar magnetosphere (and assuming 10 hours of observation).
Another interesting observable is the time dependence of the signal induced by the rotation of the NS and the plasma. Constraints from a time-domain analysis using MeerKAT observations of a nearby pulsar (PSR J2144-393) have been derived in~\cite{Battye:2023oac}, see Fig.~\ref{fig:Conversion}, as long as the possibility for improvement with the SKAO. Radio transients could also be produced by the collision of a NS with an axion DM clump, see~\cite{Walters:2024vaw}. 
As mentioned before, one can also exploit the cumulative emission from a large population of NSs, such as the one expected in the inner pc of our galaxy. Following this idea, and exploiting Green Bank Telescope data, interesting constraints have been derived in~\cite{Foster:2022fxn}, see Fig.~\ref{fig:Conversion}. On the other hand, uncertainties on the galactic center NSs population are significant~\citep{Bhura:2024jjt}.

Remarkably, an abundant population of ALPs can be directly created in pulsars, 
in particular around the pulsar polar caps~\citep{Prabhu:2021zve}, localized regions in the NS magnetosphere with an un-screened electric field ($\vec E$) such that $\vec{E} \cdot \vec{B} \neq 0$.
This phenomenon is due to the fact that $\vec{E} \cdot \vec{B}$ acts as a source term in the ALP equation of motion.
Therefore, ALPs are generically produced, regardless of whether they contribute to the DM. 
This mechanism is relatively new, and its phenomenological consequences are still being explored. 
A broadband radio signal, as opposed to the narrow lines discussed above, is generated by the conversion of the sourced relativistic ALPs while they move away from the NS.
Comparing the predicted flux from a numerical simulation and observations of 27 nearby pulsars in the frequency range 25 MHz - 8.6 GHz,
\cite{Noordhuis:2022ljw} has derived competitive limits on the axion-photon coupling, see Fig.~\ref{fig:Conversion}.
Furthermore, a fraction of the produced ALPs can also be gravitationally captured by the neutron stars, instead of  escaping. Accumulating over large timescales, they can form dense axion clouds, and lead to several striking signatures, such as a quasi-periodic cancellation
of pulsar radio pulses, and narrow radio lines, offering exciting prospects for detection with current radio telescopes and the SKAO~\citep{Noordhuis:2023wid,Caputo:2023cpv}.

In conclusion, the extreme environment around NS constitutes an ideal target to search for radio signals from ALP conversion. 
The phenomenology is quite rich and it is still under exploration. Current radio observations have already established leading limits, and future observations with the SKAO hold significant potential for ALP detection.
In particular, we can see from both Figs.~\ref{fig:ALPdecay} and \ref{fig:Conversion}, that the mass range $m_a\lesssim\mu{\rm eV}$ is of difficult reach for haloscopes whilst an ideal ground for the SKA-Low telescope.

\subsection{Birefringence} \label{sec:ALPbir}

If ALPs exist, their interaction
with photons causes the left- and the right-circularly polarized light to
travel at different velocities in the ALP field --- a phenomenon called \textit{birefringence} \citep{harar92, carro98}.
Consequently, the plane of polarization of linearly polarized light is rotated with respect to the plane at emission by $\Delta\theta_a$. The birefringence angle $\Delta\theta_a$ only depends on the axion-photon coupling $g_{a\gamma}$ and the strength of the ALP field at photon emission and detection locations.
Polarized emission from quasars observed via strong gravitational lenses, parsec-scale jets in active galactic nuclei (AGN),
and protoplanetary disks, could be effectively used to search for ALPs.
An interesting property of the ALP field is that its strength
oscillates in time 
with period ($T_a$) given by the mass as $T_a = 2\pi/m_a$. Therefore, 
$\Delta\theta_a$ also oscillates allowing us to measure $m_a$. 
For $m_a = \mathcal{O}(10^{-22}\, \rm eV)$, $T_a$ is expected to be of the
order of several months. 
Such deterministic time-variations represent a smoking gun for ALP detection.
Note also that the birefringence signal does not require the modeling of the astrophysical electromagnetic field as instead in the cases of conversion and stimulated decay.

Measurement of the birefringence angle using polarized emission at cm-wavelengths is complicated by the additional contribution from chromatic birefringence introduced by the Faraday effect when linearly polarized light propagates through a magnetized plasma. In order to mitigate the Faraday rotation, polarization angle (PA) measurements over broad bandwidths are needed because at few GHz-frequencies the Stokes\,$Q,U$ parameters vary in a complicated way due to turbulence in magnetized media \citep{sokol98}. Correction for Faraday rotation can be done by applying the technique of Stokes\,$Q,U$ fitting \citep[e.g.,][]{sulli12} or RM-synthesis \citep{brentjens2005} to broadband spectro-polarimetric observations.

{\bf Strong lensing.} \cite{basu2021} developed a new ALP detection method which makes use of the fact that strong gravitational
lensing allows simultaneous observations of multiple images from a polarized quasar, but
separated in emission time because of the lensing time delay. This crucial fact allows us to
probe the differential birefringence of time-separated images and provides an
extremely clean constraint on 
$m_a$ and $g_{a\gamma}$.
ALP-photon interaction in the curved space-time of the lens does not affect intensities \citep{Schwarz:2020jjh}. Thus, simply taking the difference of the polarization angles between the lensed images alleviates both observational
and astrophysical systematics, leaving behind a clean contribution from ALPs that allows us to achieve robustness similar to lab-based experiments. 
The differential birefringence angle ($\Delta\theta_{a,{\rm lens}}$) is given as \citep{basu2021},
\begin{equation}
    \Delta\,\theta_{a,{\rm lens}} = K \sin\!\left[\frac{m_a \Delta t}{2} \right] \sin\left(m_a t_
{\rm em} + \delta_{\rm em} - \frac{\pi}{2} \right),~K = 10^\circ\!  \left[ \frac{\rho_{a,{\rm em}} }{20\,{\rm GeV/cm^{3}} }\right]^{1/2}\!\!\! 
    \frac{g_{a\gamma}}{10^{-12}\,{\rm GeV}^{-1}} 
    \frac{10^{-22}\,{\rm eV}}{m_a}.
\label{eq:Dlens}
\end{equation}
It is important to note that, $\Delta\theta_{a, {\rm lens}}$ is discernibly large, oscillates with the same period as that of the ALP field, and, depends only on parameters of the ALP field at the emitting region denoted with the subscript `em', except for a dependence on the gravitational time delay ($\Delta t$). 
Spatial fluctuations are contained in the density ($\rho_{a,{\rm em}}$) of the ALP and the random phase ($\delta_{\rm em}$).

\begin{figure*}[t]
\vspace{-23pt}
\centering
\begin{tabular}{c}
 \includegraphics[width=0.8\columnwidth]
{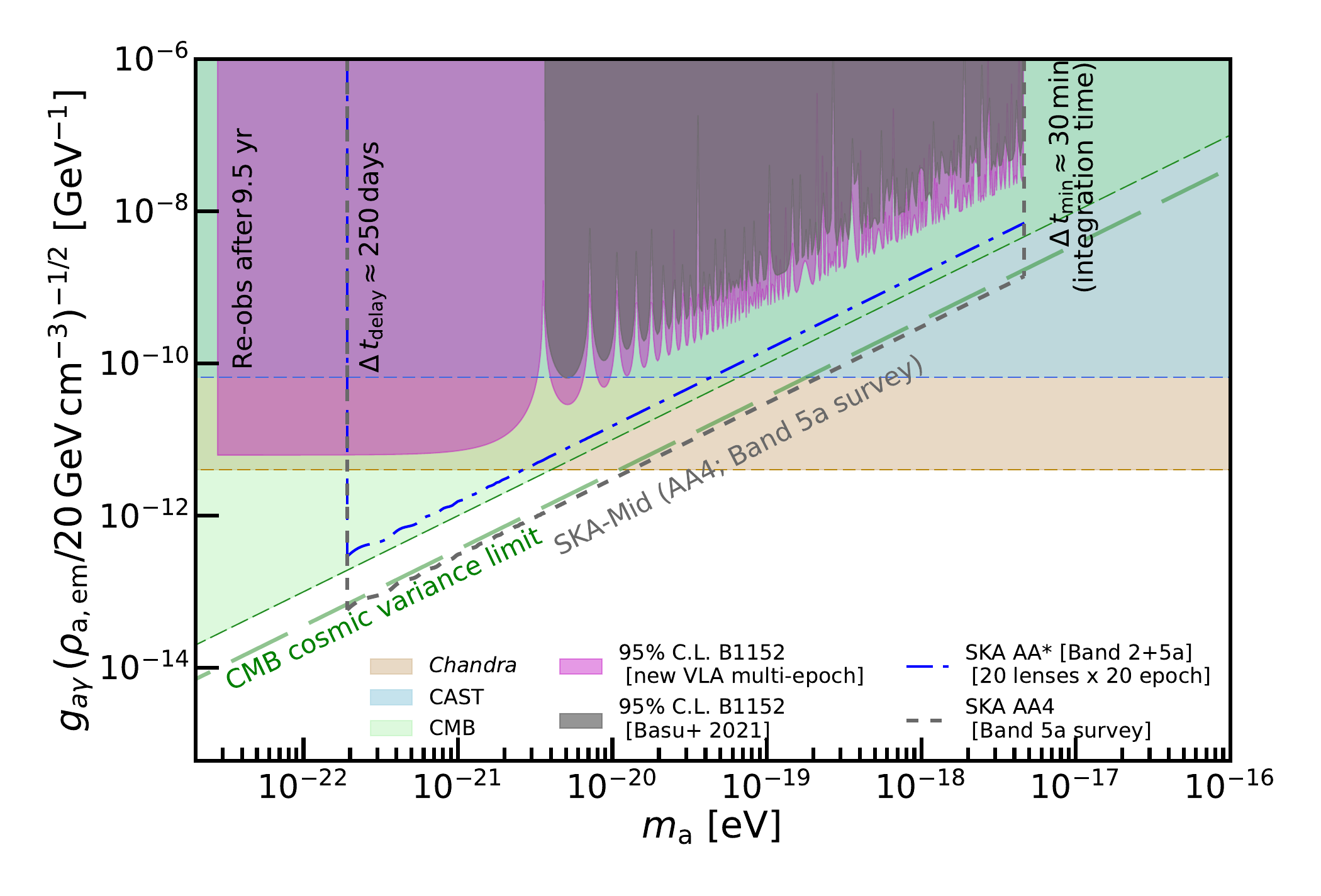} \\
\end{tabular}
\vspace{-15pt}
\caption{Exclusion region for $g_{a\gamma}$ and $m_a$ at 95\% confidence obtained by using a single epoch observations of B1152+199 (grey) and by combining data over 5 epochs \citep[pink;][]{Deshmukh:2026}. The blue dash-dotted line shows the expected parameter space than can be probed with the AA*-array by combining data from 20 selected lens systems observed over 20 epochs in Band\,2+5a. The gray dashed line shows the expected sensitivity by statistical combination of lens systems detected in Band\,5a survey with AA4-array.}
\label{fig:lensingbounds}
\vspace{-15pt}
\end{figure*}

The ALP parameter space that can be covered using strong lensing is determined by -- (i) the timescale associated with the observations (maximum $m_a$) and the time delay (minimum $m_a$), i.e., determined by the sample and observations setup; and (ii) the accuracy of $\Delta\theta_{a,{\rm lens}}$, which sets the bound on $g_{a\gamma}$ \citep{basu2021}. The accuracy of $\Delta\theta_{a,{\rm lens}}$ 
is mainly governed by the error associated with measuring the Faraday rotation measure (RM) when correcting the PA. High accuracy requires (1) precise measurement of RM at low frequencies, e.g., Band\,2 or below \citep{brentjens2005}, and/or (2) improving the signal-to-noise ratio through statistical combination of a large sample of lens systems \citep{basu2021}. The precision in $\Delta\theta_{a,{\rm lens}}$ using (1) is limited by the $\approx0.3\textrm{--}0.8^{\prime\prime}$ resolution of the SKA-Mid telescope in Band\,2, limiting to lens separation $\gtrsim1^{\prime\prime}$, and therefore the required precision can be achieved through multi-epoch observations of a selected sample of lens systems. In Fig.~\ref{fig:lensingbounds} we show the parameter space that can be probed through 20 epoch observations of 20 lens systems using the AA* and/or AA4 array layout of SKA-Mid as the blue dash-dotted line. This will significantly improve upon the current best robust bound provided by helioscopes \citep[][]{CAST2017}. For (2), statistical combination of lens systems detected in spectropolarimetric surveys at Band\,5a (or above) can also provide strong constraints on ultralight-ALPs. With the higher resolution ($\sim0.05$\,arcsec) at Band\,5a with the AA4 layout, $\mathcal{O}(10^5)$ lens systems are expected to be detected~\citep{Pandey-Pommier02.2026.SKA}. Even the pessimistic assumption of 5\% of polarized sources allows to reach the level of the (cosmic variance limited) CMB-bound \citep{fedde19}, shown as the gray dashed line in Fig.~\ref{fig:lensingbounds}. These predictions are based on an extrapolation from \cite{basu2021} (grey shaded region), and new constraints obtained from 5-epoch combination of VLA data from \cite{Deshmukh:2026} (pink shaded region).

\subsection{Synergies with other indirect searches }
Multi-wavelength observations can contribute to constrain the axion-photon coupling.
Optical and UV polarization measurements of thermal radiation from magnetic white dwarf can probe a linear polarization induced from photon-axion conversion. ALP production and conversion in the Sun can produce an X-ray emission. The energy spectra of extra-galactic gamma-ray and X-ray sources get modified by the photon-axion conversion. Microwave CMB experiments are sensitive to the cosmic birefringence effect. For an extensive list of works exploring the ALP and DP parameter space at different frequencies with astrophysical and cosmological observations, see the repository~\href{https://cajohare.github.io/AxionLimits/}{https://cajohare.github.io/AxionLimits/}. The reach level is depicted in Fig.~\ref{fig:Conversion} and \ref{fig:lensingbounds}.

Laboratory experiments include axion helioscopes, that look for ALPs coming from the Sun (dark red bounds in Figs.~\ref{fig:ALPdecay} and \ref{fig:Conversion}), and haloscopes (light red) probing the DM ALPs or DPs at our location.

As it can be understood by comparing our forecasts with other techniques in Figs.~\ref{fig:ALPdecay}, \ref{fig:Conversion}, and \ref{fig:lensingbounds}, observations with the SKA telescopes can be very competitive to constrain the ALP and DP parameter spaces.


\section{Conclusion}
The construction of the SKAO has been aiming to address several outstanding scientific cases. One of the most pressing scientific challenges of our time concerns understanding the fundamental nature of DM.
In this Chapter, we outlined how the SKAO can investigate compelling classes of particle DM candidates across 30 orders of magnitude in mass: WIMPs, ALPs and dark photons. Forecasts for the SKA-Low and SKA-Mid telescopes in the AA4 baseline design are presented. 

The observational strategies include a variety of observations (continuum, spectral, polarimetric, transient), and targets, as summarized in Table~\ref{tab:summary}.
There are different types of signatures we are looking for, as sketched in the cartoon in Fig.~\ref{fig:cartoon}.

To mention a few relevant examples, we have reported that observing:\\ - synchrotron radiation in a selection of dwarf galaxies and galaxy clusters, for a few tens of hours in continuum mode with SKA-Low and short-baselines of SKA-Mid AA4 across different frequency bands, can provide the best bound on the WIMP annihilation cross section from indirect searches, constraining the reference thermal value up to at least 1 TeV in mass (see Figs.~\ref{fig:clusterz02}, \ref{fig:ret2-ska} and \ref{fig:summaryWIMP});\\ - ALP decay from the Galactic anti-center with SKA-Low AA4 in autocorrelation mode for 100 hours with $10^4$ spectral channels can improve current laboratory bounds on the axion-photon coupling for ALP masses around $10^{-6}$ eV (see Fig.~\ref{fig:ALPdecay});\\ - ALP conversion in isolated magnetars with SKA-Mid AA4 (in the Sun with SKA-Low AA4) for 10 (100) hours with high spectral resolution can tighten the bound around $10^{-5}$ eV ($10^{-7}$ eV), with a significant impact also for the dark photon (see Fig.~\ref{fig:Conversion});\\ - birefringence from lens systems in spectropolarimetric surveys at Band 5a with AA4 can improve ALP laboratory bounds at very low masses up to $10^{-19}$ eV (see Fig.~\ref{fig:lensingbounds}). \\
On top of this list, other relevant techniques are discussed in the above Sections.

This variety makes the scientific case of particle DM robust against different telescope setups. At the same time, it highlights the importance of commensality, and exploitation of synergies with different Science Working Groups.

The comprehensive program described in this Chapter can make SKAO play an important role in the DM quest.
\section*{Contribution}
Each of the authors contributed to the writing of specific sections and to the review of the complete manuscript.
The work was coordinated by MR.
\section*{Acknowledgments}
MR and MT acknowledge support from the  Research grant TAsP (Theoretical Astroparticle Physics) funded by \textsc{infn}. The work of MR is supported by by the European Union – Next Generation EU and by the Italian Ministry of University and Research (MUR) via the PRIN 2022 Project No. 20228WHTYC – CUP: D53C24003550006. MT acknowledges the research grant “Addressing systematic uncertainties in searches for dark matter No. 2022F2843” funded by MIUR.

AB and DJS acknowledge support by the Bundesministerium für Forschung, Technologie und Raumfahrt (BMFTR) under ErUM-Pro grants 05A17PB1, 05A20PBA, and 05A23PBA.

E. Todarello has received funding from the STFC Consolidated Grant [ST/T000732/1] and is supported by the European Union’s Horizon 2020
research and innovation programme under the Marie Skłodowska-Curie grant
agreement No 101204903 (APARAX).

CU is supported by the European Union (ERC StG, LSS\_BeyondAverage, 101075919).

E. Tolley acknowledges support from the Swiss National Science Foundation under the SNSF Starting Grant ``Deep Waves'' (218396).

\bibliographystyle{abbrvnat-maxbibnames4}
\bibliography{chapter} 

\end{document}